# Comparison of Matlantis and VASP bulk formation and surface energies in metal hydrides, carbides, nitrides, oxides, and sulfides

Shinya Mine[1], Takashi Toyao[1], Ken-ichi Shimizu[1], and Yoyo Hinuma[2*]

[1] Institute for Catalysis, Hokkaido University, N-21, W-10, 1-5, Sapporo 001-0021, Japan
[2] Department of Energy and Environment, National Institute of Advanced Industrial Science and Technology (AIST), 1-8-31, Midorigaoka, Ikeda 563-8577, Japan

* y.hinuma@aist.go.jp

**Abstract**

Generic neural network potentials without forcing users to train potentials could result in significantly acceleration of total energy calculations. Takamoto *et al.* [Nat. Commun. (2022), 13, 2991] developed such a deep neural network potential (NNP) and made it available in their Matlantis package. We compared the Matlantis bulk formation, surface, and surface O vacancy formation energies of metal hydrides, carbides, nitrides, oxides, and sulfides with our previously calculated VASP values obtained from first-principles with the PBEsol(+$U$) functional. Matlantis bulk formation energies were consistently ~0.1 eV/atom larger and the surface energies were typically ~10 meV/Å$^2$ smaller than the VASP counterpart. Surface O vacancy formation energies were generally underestimated within ~0.8 eV. These results suggest that Matlantis energies could serve as a relatively good descriptor of the VASP bulk formation and surface energies.

## 1. Introduction

Analyzing results of massively large-scale calculations is one major direction of materials informatics. Using energy calculators with acceptable accuracy that are computationally much less expensive than density functional theory (DFT)-based first-principles calculations [1-2] can increase the search space considerably. Neural network potentials (NNPs)[3-6] go beyond classic empirical potentials regarding the capability to handle complicated interactions. A descriptor, in materials informatics jargon, is a low-cost quantity that reflects high-cost quantities that are difficult to obtain. A decent NNP serves as a descriptor of the total energy and also coarsely relax atom positions. As a result, use of NNPs can therefore reduce the time running first-principles calculations to initially relax atoms from very favorable positions. A perfect NNP cannot make standard first-principles calculations totally obsolete. DFT is backed up with theories in quantum mechanics. On the other hand, the NNP is a black box trained to estimate total energies and forces accurately and intrinsically cannot derive accurate energy levels, thus important quantities such as the band gap, defect levels, and ionization potential cannot be obtained.

Takamoto *et al.* [7] recently developed a deep neural network potential, PreFerred Potential (PFP), and made it available in their Matlantis package. They claim that their NNP is universal, meaning that there is no need to fine-tune potentials for individual applications, supports almost all naturally occurring elements, and using the NNP results in $\sim 10^4$ times faster calculations than standard first-principles calculations while maintaining reasonable accuracy.

This study evaluated the prediction performance of bulk formation energy and surface energy in 3, 4, and 5 hydrides, carbides, nitrides, oxides, and sulfides discussed in Ref. [8] (groups 3-5 HCNOS-ides set) and $d^0$ and $d^{10}$ binary oxides with cation valences between +1 and +6 (binary oxides set, Ref. [9]).

2. **Computational details**

Bulk formation, surface, and surface O vacancy formation energies from Matlantis and VASP [10-11] codes were compared to check how Matlantis energies are useful as descriptors of VASP energies. Matlantis energies were obtained using the PFP model v3.0.0 [7] based on the generalized gradient approximation (GGA) [12] and the Perdew-Burke-Ernzerhof (PBE) functional [13]. All-atom energy optimizations were conducted using the Broyden–Fletcher–Goldfarb–Shanno (BFGS) algorithm with a force threshold of 0.01 eV Å$^{-1}$. Structural optimization of the bulk model preserved symmetry and relaxed cell shape and atomic positions. On the other hand, the structural optimization of the surface slab model preserved the symmetry and relaxed the atomic positions while keeping the cell shape fixed.

VASP energies are from the PBEsol functional [14] as used in our previous research. [8-9] No *U* was used in the groups 3-5 HCNOS-ides set because most hydrides, carbides and nitrides were metallic and therefore use of *U* is not suitable. In contrast, the Hubbard *U* correction [15-16] was taken into account in the binary oxides set. Only one choice of the effective $U$ value is allowed in Matlantis, which is 3.25 eV in V, 4.38 eV in Mo, and 4.0 eV in Cu for valence $d$ states. On the other hand, the effective $U$ value in VASP calculations was 3 eV for the valence $d$ states of La, Ti, Zr, Hf, V, Rh, Au, and Hg, 3.5 eV for Mo, and 5 eV for the valence $d$ states of Cu, Zn, and Cd. Additionally, 5 eV was used for 4$f$ states of Ce and Gd. No dispersion (van der Waals) corrections were considered. Spin polarized calculations were conducted with ferromagnetic initial spin.

## 3. Results and discussion

Bulk formation energies are compared in **Figure 1**. Information on the systems considered in group 3, 4, and 5 hydrides, carbides, nitrides, oxides, and sulfides (groups 3-5 HCNOS-ides set, **Figures 1(a-e)**), respectively, are shown in Tables 1-5 in Ref. [8], respectively, while those of the binary oxides set (**Figure 1(f)**) is given in Tables 1 and 2 of Ref. [9]. The Matlantis bulk formation energy, $E_{BM}$, is roughly 0.1 eV/atom larger (dotted line) than the corresponding VASP energy. The average and standard deviation of $E_{BM}$-$E_{BV}$ is (0.10, 0.03), (0.06, 0.04), (0.16, 0.04), (0.08, 0.06), and (0.08, 0.04) in hydrides, carbides, nitrides, oxides, and sulfides in the groups 3-5 HCNOS-ides set and (0.10, 0.08) in the binary oxides set (units in eV/atom).

**Figure 2** shows the relation between VASP and Matlantis surface energies. In general, the lowest surface energy termination among bulk cleaved with various orientations was chosen for each compound, although multiple low energy terminations were considered in some systems. Results for group 3, 4, and 5 hydrides, carbides, nitrides, oxides, and sulfides (groups 3-5 HCNOS-ides set) are shown in **Figures 2(a-e)**, respectively. Slab geometries and surface energies are from "thick" slabs in Tables S1-S5 of Ref. [9], and surface terminations are given in Figures 1 and 2 of Ref. [9]. For the binary oxides set in **Figure 2(f)**, slab geometries and VASP energies are from "thin" slabs in Table SI-1 and SI-2 of Ref. [9], respectively, and surface terminations are given in Supplementary Figure SI-1 of Ref. [9].

The Matlantis surface energy, $E_{SM}$ is, typically approximately 10 meV/Å$^2$ smaller than the VASP surface energy, $E_{SV}$ (center line in **Figure 2**). In most cases points lie between the top and bottom lines in **Figure 2**, satisfying -20 meV/Å$^2$ < $E_{SM}$ - $E_{SV}$ < 0 meV/Å$^2$. Oxides are mostly in a narrower range of -10 meV/Å$^2$ < $E_{SM}$ - $E_{SV}$ < 0 meV/Å$^2$. This result suggests that ($E_{SM}$ - 10 meV/Å$^2$) serves as a reasonable approximation of $E_{SV}$ with an "error bar" of about ±10 meV/Å$^2$. The almost constant shift could arise from the difference in the underlying functional: PBE (+$U$) was used when learning the PFP model while PBEsol(+$U$) was used in our past VASP calculations because it provided reasonable energetics and crystal structures in our previous work on binary oxides with formally closed-shell electronic structures.[17]

The VASP surface O vacancy formation energies ($E_{Ovac}$) of the binary oxides set, as calculated in Ref. [9], are compared with Matlantis values in **Figure 3(a)**. Matlantis values are almost the same or underestimated compared to VASP values, and overestimation, when it happens, is small. Compounds where the difference between VASP and Matlantis is over 0.8 eV are labeled in **Figure 3(a)**. Many are early transition metal oxides that are $d^0$ when defect-free but some atoms have valence $d$-electrons upon O vacancy formation. The value of $U$ can become relevant in such cases, and considering $U$ in VASP and not in Matlantis is one cause of the energy difference. However, some oxides with cations where $U$ is not imposed have large discrepancies. The Matlantis $E_{Ovac}$ of GeO$_2$ underestimates the VASP $E_{Ovac}$ by 1.5 eV. The bonding environment of the O immediately under the removed O is different in the two approaches; its distance from the subsurface cation differs between VASP and Matlantis (**Figures 3(b,c)**, respectively). Favoring different coordination environments means that the bond strengths near undercoordinated are evaluated differently in VASP and Matlantis. The discrepancy in BaO is 1.1 eV. The four Ba in the topmost layer next to the O vacancy move away further from the O vacancy in Matlantis compared to VASP. Such differences in bond strength evaluation leads to different $E_{Ovac}$ values with the two approaches.

## 4. Conclusions

Matlantis estimates VASP energies relatively well. The Matlantis bulk formation energy is consistently ~0.1 eV/atom larger and the surface energy is typically ~10 meV/Å$^2$ smaller than the VASP counterpart. Surface O vacancy formation energies are generally underestimated within ~0.8 eV. These results suggest that Matlantis energies could serve as a relatively good descriptor of the VASP bulk formation and surface energies.

**Disclosure statement**
The authors have no potential conflict of interest.

**Additional information**
**Funding**

This research was supported by the JST-CREST project JPMJCR17J3, KAKENHI (Grant Nos. 21H05101 and 22K14538), and the Joint Usage/Research Center for Catalysis (proposals 22AY0027 and 23AY0192).

**Miscellaneous**

Information on Matlantis is available at https://matlantis.com/ . The VESTA code [18] was used to draw **Figures 3(b,c)**.

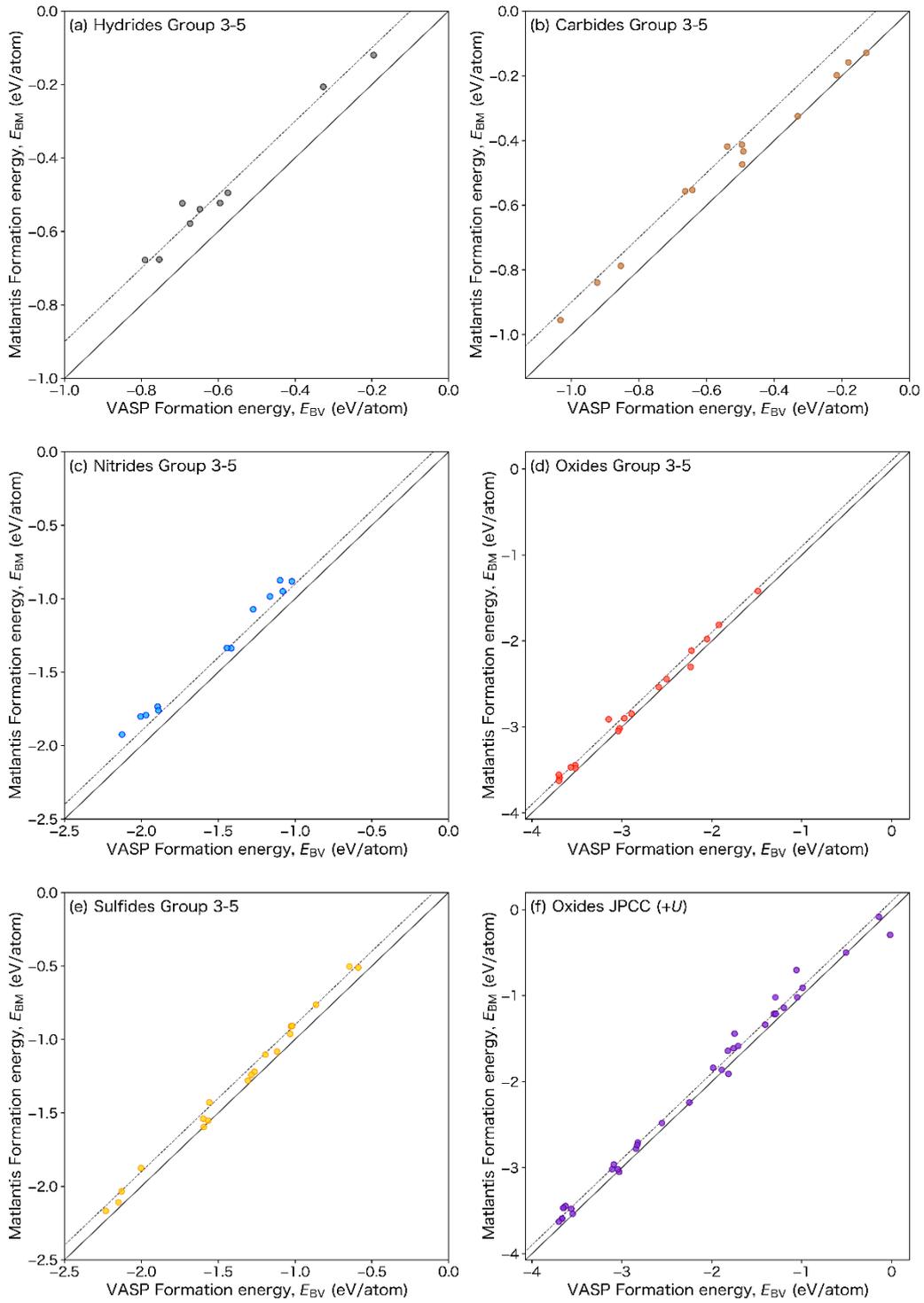

Figure. 1. Relation between Matlantis and VASP formation energies for groups 3-5 (a) hydrides, (b) carbides, (c) nitrides, (d) oxides, and (e) sulfides in Ref. [8] and (f) $d^0$ and $d^{10}$ binary oxides in Ref. [9]. The thin and thick solid lines are $y=x+0.1$ and $y=x$, respectively.

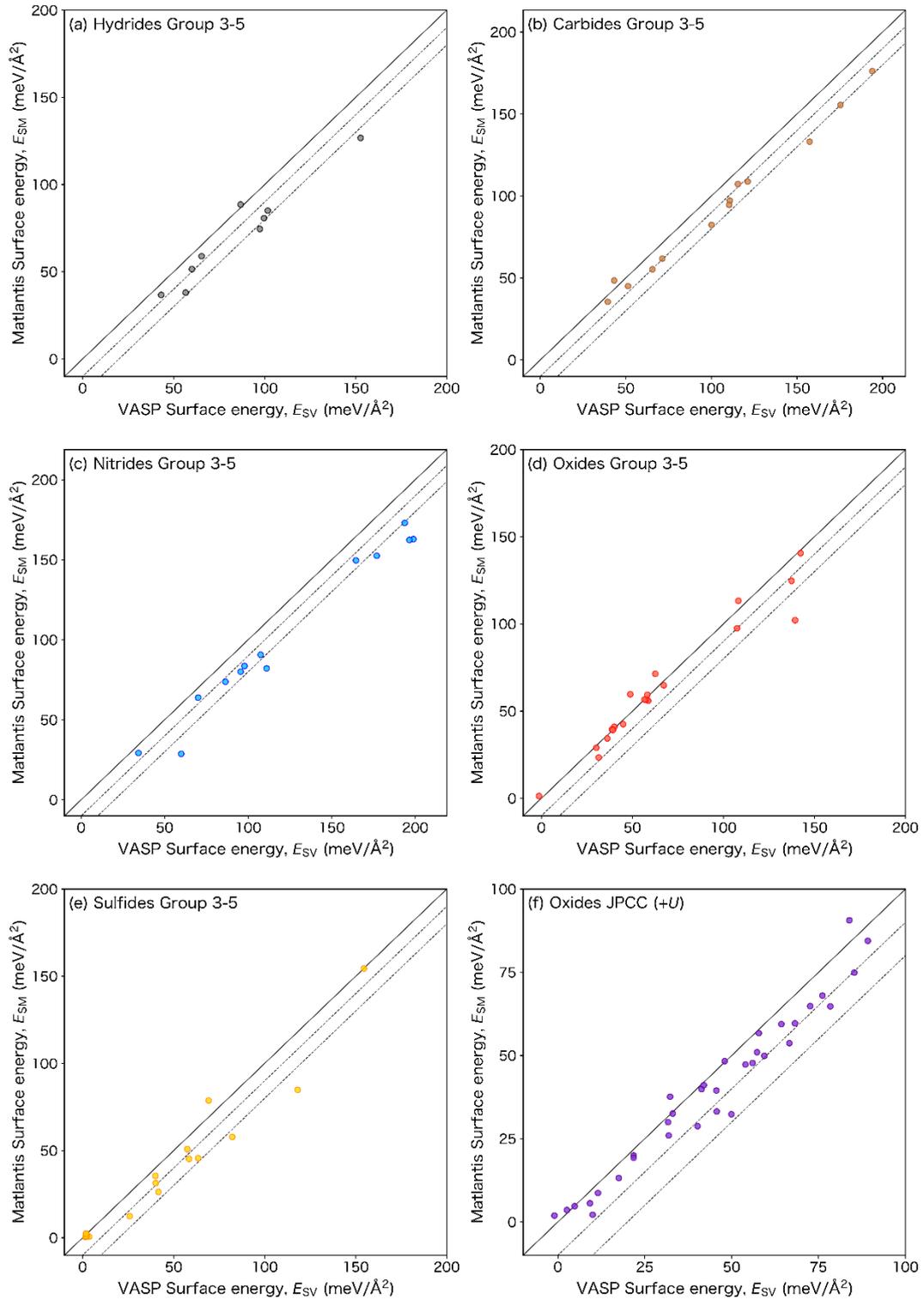

Figure. 2. Relation between Matlantis and VASP formation energies for groups 3-5 (a) hydrides, (b) carbides, (c) nitrides, (d) oxides, and (e) sulfides in Ref. [8] and (f) $d^0$ and $d^{10}$ binary oxides in Ref. [9]. The thick line is $y=x$. Dotted lines $y=x$-10 and $y=x$-20 are additionally shown as thin lines.

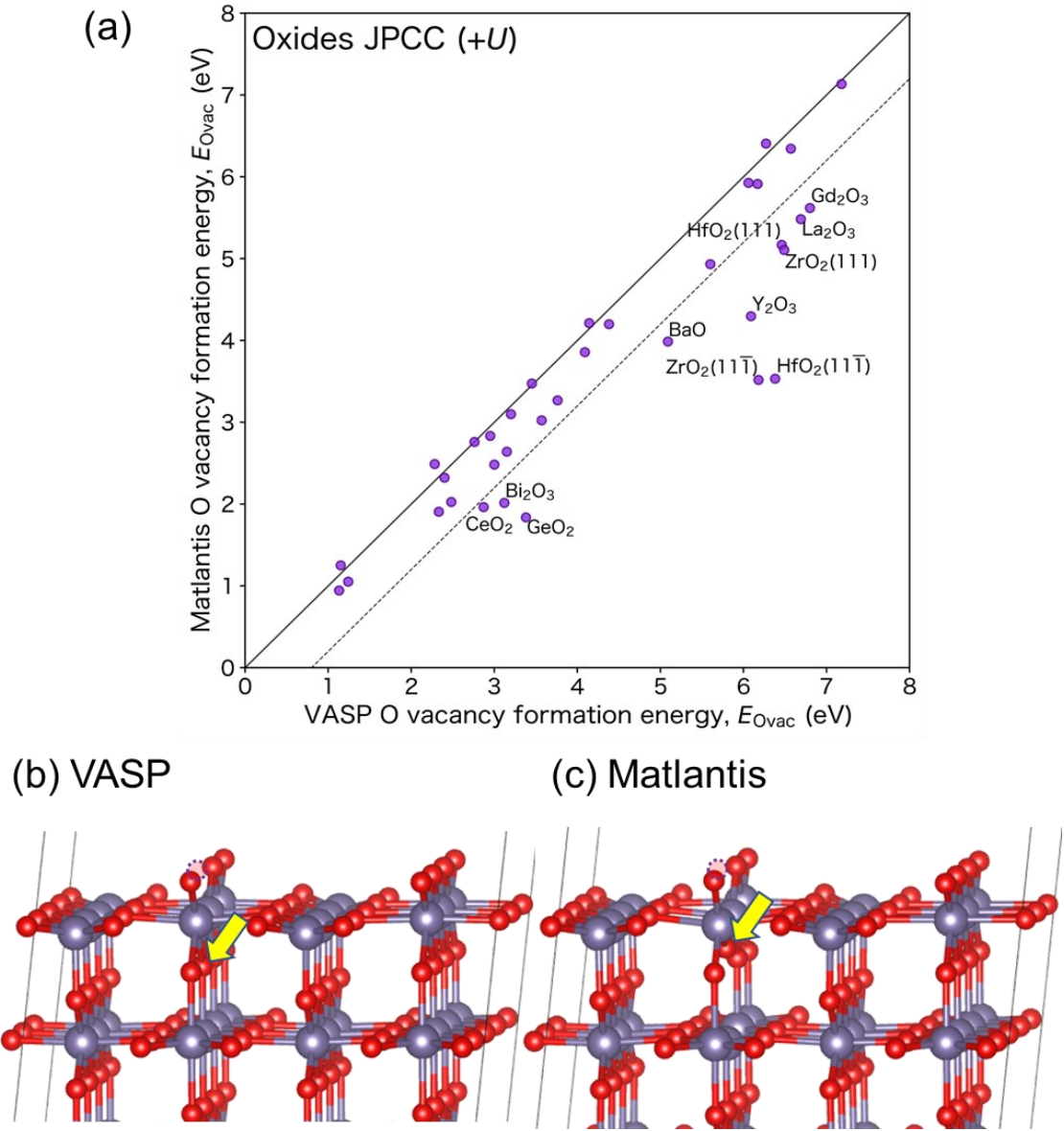

Figure. 3. (a) Comparison of Matlantis and VASP $E_{Ovac}$ values for the binary oxides set. The line $y=x-0.8$ is shown as a dotted line. Comparison of (b) VASP and (c) Matlantis relaxed surfaces of $GeO_2$ (110) when there is an O vacancy at the surface (pink circle). Note the position of O at the yellow arrow, which is immediately below the O vacancy. Purple and red balls indicate Ge and O, respectively.

9